\documentclass{aa}
\usepackage{bm}
\usepackage{showlabels}[1]
\usepackage{lscape,rotating}
\usepackage{natbib}
\usepackage{hyperref}
\usepackage{amsmath,amssymb}
\usepackage{times}
\usepackage{graphicx}
\usepackage{fancyhdr}
\usepackage{url}
\graphicspath{{./fig/}{./png/}}
\def \Sh  {\mbox{Sh}}

\def \Rm  {\mbox{Re}_{\rm M}}
\def \Pm  {\mbox{Pr}_{\rm M}}
\def \cs  {c_{\rm s}}
\def \kf  {k_{\rm f}}
\def \urms{u_{\rm rms}}
\def \etat{\eta_{\rm t}}
\newcommand{\SSSS}{\mbox{\boldmath ${\sf S}$} {}}
\newcommand{\meanEMF}{\overline{\mbox{\boldmath ${\cal E}$}}{}}{}
\newcommand{\meanBB}{\overline{\bm{B}}}
\newcommand{\meanJJ}{\overline{\bm{J}}}
\newcommand{\meanWW}{\overline{\bm{W}}}
\newcommand{\meanB}{\overline{B}}
\def\onehalf{{\textstyle{1\over2}}}

\title{Alpha effect and diffusivity in helical turbulence with shear}
\authorrunning{D. Mitra et al.} 
\author{Dhrubaditya Mitra\inst{1}, Petri J. K\"apyl\"a\inst{2}, 
Reza Tavakol\inst{1}, and Axel Brandenburg\inst{3} } 
\institute{
Astronomy unit, School of Mathematical Sciences, Queen Mary University of London, 
Mile End Road, London E1 4NS, UK
\and
Observatory, T\"ahtitorninm\"aki (PO Box 14), FI-00014,
University of Helsinki, Finland
\and
NORDITA, Roslagstullsbacken 23, SE-10691 Stockholm, Sweden
}
\date{\today,~ $ $Revision: 1.222 $ $}
\begin{document}
\abstract
{}
{We study the dependence of turbulent transport coefficients,
such as the components of the $\alpha$ tensor ($\alpha_{ij}$)
and the turbulent magnetic diffusivity tensor ($\eta_{ij}$), on shear
and magnetic Reynolds number in the presence of helical forcing.}
{We use three-dimensional direct numerical simulations
with periodic boundary conditions and measure the turbulent transport 
coefficients using the kinematic test field method.
In all cases the magnetic Prandtl number is taken as unity.}
{We find that with increasing shear the diagonal components 
of $\alpha_{ij}$ quench, whereas those of $\eta_{ij}$ increase. 
The antisymmetric parts of both tensors increase with increasing shear.
We also propose a simple expression for the turbulent pumping velocity
(or $\gamma$ effect). This pumping velocity is proportional
to the kinetic helicity of the turbulence and the vorticity of the mean flow.
For negative helicity, i.e.\ for a positive trace of $\alpha_{ij}$,
it points in the direction of
the mean vorticity, i.e.\ perpendicular to the plane
of the shear flow. Our simulations support this expression for 
low shear and magnetic Reynolds number.
The transport coefficients depend on the wavenumber of the
mean flow in a Lorentzian fashion, just as for non-shearing
turbulence.
}
{}
\keywords{magnetohydrodynamics (MHD) -- hydrodynamics -- turbulence -- magnetic fields}
\maketitle
\section{Introduction}
Understanding the origin of non-uniform large-scale 
magnetic fields in stars, galaxies, and accretion discs continues 
to pose important challenges. Such fields are commonly 
thought to be the result of dynamo action that converts the kinetic
energy of turbulent motions and large-scale shear into magnetic energy.
The usual framework for studying 
such dynamo actions is mean field electrodynamics \citep[e.g.][]{KR80}.
Over the years, however, the applicability of this framework has been
questioned \citep{P81,VC92}. 
In particular, an important debate in this connection revolves
around the role played by magnetic helicity \citep{GD94}.
Magnetic helicity is conserved for ideal 
(non-dissipative) magnetohydrodynamics (MHD) 
and also in the limit of large magnetic Reynolds numbers.
This conservation places severe 
constraints on the growth of the mean magnetic field and may   
regulate the quenching of the $\alpha$ effect
as the magnetic Reynolds number ($\Rm$) increases~\citep{BS05}.
Under certain circumstances (no magnetic helicity fluxes), $\alpha$ quenching
becomes more severe with $\alpha$ proportional to $\Rm^{-1}$.
It has been suggested that this can be alleviated by 
expelling magnetic helicity through open boundaries \citep{BF00,KMRS00},
possibly mediated by shear \citep{VC01,SB04,BS04}.
Furthermore, shear itself is an important ingredient 
in MHD dynamos in solar and stellar settings as, for example,
in the solar tachocline. Hence, it is important to understand 
how shear affects the turbulent transport coefficients, in particular
the components of the $\alpha_{ij}$ and $\eta_{ij}$ tensors.

Several studies have recently looked at various 
aspects of this problem \citep[see, e.g.,][]{RK03,RK04,Bra05,RK06,
RS06,LK08,BRRK08}. These works employ different tools and make different 
assumptions and are often applicable to limited regions of the parameter space.
As a consequence, care must be taken in comparing these results.
For example, using semi-analytical tools, which treat the 
nonlinear Lorentz force feedback perturbatively, \cite{LK08} have found that 
shear can reduce  $\alpha$ in helically forced turbulence.  
This is analogous to the ``rotational quenching'' of turbulent transport
coefficients with increasing Coriolis or inverse Rossby numbers
\citep{KRP94,PRK96}.
There are also other related cases in which the presence of
shear enhances the growth rate of the dynamo. For example; 
using direct numerical simulations of 
the MHD equations in the presence of shear 
and {\em non-helical} forcing, \cite{YHSKRICM07,YHRSKRCM08} have found
large-scale dynamos whose growth rate increases linearly with shear. 
Such scaling has also been found for $\alpha$--shear dynamos where the
$\alpha$ effect is due to stratified convection with shear
\citep{KKB08}.
Furthermore, using the kinematic test field method 
(described below), \cite{BRRK08} have studied the dynamo 
coefficients in the presence of shear, but in the absence of helicity, 
and they find that Gaussian fluctuations of $\alpha$ about 
zero are strong enough to drive an incoherent 
$\alpha$--shear dynamo~\citep{VB97,P07}.
The significance of the incoherent $\alpha$ effect has also been stressed
by \cite{HP08}, although their system may have also had a net $\alpha$ effect.
They dismissed this on the grounds that for an imposed uniform magnetic field
$\alpha$ is very small.
However, this result disagrees with recent calculations of $\alpha$
using the test field method \citep{KKB08b}.

In this paper we use three-dimensional direct numerical simulations 
of the kinematic test field equations with {\em helical} forcing in order 
to study the dependence of turbulent transport coefficients on shear 
and magnetic Reynolds number.
In Section 2 we give a brief account of our model.
Section 3 contains our results and finally we conclude in Section 4.
\section{The model}
\label{TheModel}
We use the test field method \citep{SRSRC05,SRSRC07}
to calculate the turbulent transport coefficients.
This method and its modification in the presence of large-scale shear
are described in \citep{BRRK08}.
Here we just point out the essence of the method and elaborate only on
those aspects where our treatment differs from their paper.

In the presence of large-scale shear, the equations of magnetohydrodynamics
are treated in the following way. Writing the velocity as the 
sum ${\bm U} + {\bm U}^{S}$, where the large-scale shear
velocity ${\bm U}^{S}=Sx\bm{\hat{y}}$ with a constant shear $S$,
and assuming an isothermal equation of state characterised by the sound 
speed $c_{\rm s}$, the momentum equation becomes 
\begin{equation}
\frac{{\mathcal D} {\bm U}}{{\mathcal D} t} = 
    - {\bm U} \cdot \bm{\nabla} {\bm U} -
S U_x \bm{\hat{y}} - \cs^2\bm{\nabla} \ln \rho + {\bm f} + {\bm F}_{{\rm visc}}.
\label{duu_dt}
\end{equation}
Here ${\bm F}_{{\rm visc}} = \rho^{-1}\bm{\nabla}\cdot(2\rho\nu{\SSSS})$
is the viscous force, ${\sf S}_{ij}= \frac{1}{2}(U_{i,j}+U_{j,i}) -
     \frac{1}{3}\delta_{ij}\bm{\nabla} \cdot {\bm U}$ 
is the traceless rate of strain tensor
(not to be confused with the shear parameter $S$),
$\nu$ is the kinematic viscosity,
$\rho$ is the fluid density, $\cs$ is the isothermal sound speed and  
\begin{equation}
\frac{{\cal D} }{{\cal D} t} \equiv \frac{\partial}{\partial t} + 
              Sx\frac{\partial}{\partial y}.
\end{equation}
As our external forcing ${\bm f}$ we employ
helical, white-in-time, random forcing described in \citep{B01}. 
In this paper we consider the purely kinematic
problem, so there is no feedback in
Eq.~(\ref{duu_dt}) due to the Lorentz force.
In addition to the momentum equation, we have the continuity equation
\begin{equation}
\frac{{\mathcal D} \rho}{{\mathcal D} t} = - {\bm U} \cdot \bm{\nabla} \ln \rho - 
\bm{\nabla}\cdot {\bm U},
\label{dlnrho_dt}
\end{equation} 
and the uncurled induction equation in the Weyl gauge,
\begin{equation}
\frac{{\mathcal D}{\bm A}}{{\mathcal D}t}=-SA_y\bm{\hat{x}}+{\bm U}\times{\bm B}
     -\eta\mu_0{\bm J}.
\label{daa_dt}
\end{equation}
Here the magnetic field is ${\bm B} = \bm{\nabla} \times {\bm A}$,
the current density is ${\bm J} = \bm{\nabla} \times {\bm B}/\mu_0$,
and $\mu_0$ and $\eta$ are the vacuum permeability
and the molecular magnetic diffusivity, respectively.
In the mean field approach to MHD one usually decomposes the 
magnetic field (or magnetic vector potential) and velocity into 
mean (indicated by an overbar) and fluctuating parts respectively
\begin{equation}
{\bm A} = \overline{\bm A}  + {\bm a}, ~~~~ 
{\bm U} = \overline{\bm U}  + {\bm u}.
\label{mfield}
\end{equation}
The equations satisfied by the mean and fluctuating parts of the magnetic vector
potential are given by
\begin{equation}
\frac{\mathcal{D} {\overline{\bm A}}}{\mathcal{D} t}  = 
    -S{\overline A}_y\bm{\hat{x}} + {\overline{\bm U}}\times {\overline{\bm B}}
    +\meanEMF -\eta\mu_0{\overline{\bm J}},
\label{dabar_dt}
\end{equation}
and 
\begin{equation}
\frac{{\mathcal D}{\bm a}}{{\mathcal D}t}= -Sa_y\bm{\hat{x}} + 
    {\overline {\bm U}}\times{\bm b} + {\bm u}\times{\overline{\bm B}} +
    {\bm u} \times{\bm b} - \meanEMF - \eta\mu_0{\bm j},
\label{da_dt}
\end{equation}
where ${\bm j} = {\bm J} - {\overline {\bm J}}$ and 
$\meanEMF = \overline{{\bm u} \times {\bm b}}$ is the 
mean turbulent electromotive force. Specification 
of this second order quantity in terms of the mean field constitutes 
a closure problem. A common procedure is to expand $\meanEMF$ 
in terms of the mean field $\overline{\bm B}$ and its derivatives,
\begin{equation}
\overline{\mathcal{E}}_i = \overline{\mathcal{E}}_{0i} + 
                  \alpha_{ij} {\overline B_j} +
                  \eta_{ijk} {\overline B_{j,k}} \, ,
\label{emf}
\end{equation}
where $\alpha_{ij}$ and $\eta_{ijk}$ are the tensorial turbulent 
transport coefficients, and $\meanEMF_0$ quantifies additional
contributions that arise even in the absence of a mean field,
owing to small-scale dynamo action, for example.
Here and throughout, summation is assumed over repeated indices.
In the test field method we take the mean magnetic field
to be a given `test field' ${\overline {\bm B}}$, and calculate these 
tensors by measuring $\meanEMF$ by solving Eq.~(\ref{da_dt}) 
simultaneously with Eqs.~(\ref{duu_dt}) and (\ref{dlnrho_dt}), while
using Eq.~(\ref{mfield}) to find ${\bm u}$.
In order to find all the components of the $\alpha_{ij}$ and $\eta_{ij}$
tensors, and not just projections relevant to the actual fields, 
one has to use an orthogonal set of test fields
and solve Eq.~(\ref{da_dt}) for each of them. 

In our simulations we employ averages over $x$ and
$y$ directions to define our mean fields, which are therefore functions of 
$z$ and $t$ only. Thus, all other components 
of ${\overline B}_{j,k}$ except ${\overline B}_{x,z}$ and
${\overline B}_{y,z}$ vanish, and the
$\alpha$ and $\eta$ tensors can be written as rank two tensors
with $\eta_{i1} = \eta_{i23}$ and $\eta_{i2} = -\eta_{i13}$ for $i,j=(1,2)$
\citep{BRRK08}. Now consider, as an example, the following two test fields, 
\begin{eqnarray}
{\overline B}^{c1}_i = B_0(\cos kz,0,0),
&&\mu_0{\overline J}^{c1}_i = kB_0(0,-\sin kz,0), \\
{\overline B}^{s1}_i = B_0(\sin kz,0,0),
&&\mu_0{\overline J}^{s1}_i = kB_0(0,+\cos kz,0),
\label{Btest}
\end{eqnarray}
which we use to compute the two corresponding mean electromotive forces
$\meanEMF^{c1}$ and $\meanEMF^{s1}$. The relevant components of the
$\alpha$ and $\eta$ tensors are then given by 
\begin{equation}
\left( \begin{array}{c} \alpha_{i1} \\ - \eta_{i2}k \end{array} \right) =
\left( \begin{array}{cc}\;\cos kz & \sin kz \\ 
                     \!\!-\sin kz & \cos kz \end{array}\right)
\left( \begin{array}{c} {\overline {\mathcal E}}^{c1}_i \\ 
      {\overline{\mathcal E}}^{s1}_i \end{array} \right)  .
\end{equation} 
The other 2+2 components of the $\alpha$ and $\eta$ tensors can be similarly determined
by using the test fields,
\begin{equation} {\overline B}^{c2}_i = B_0(0,\cos kz,0), ~~~
\/{\overline B}^{s2}_i = kB_0(0,\sin kz,0)  .
\label{Btest2}
\end{equation} 
In what follows we denote a test field by ${\overline{\bm  B}}^{pq}$,
where $p=c,s$ and $q=1,2$. 
A particular small-scale magnetic vector potential
that develops in response to the ${\overline {\bm B}}^{pq}$ is denoted by 
${\bm a}^{pq}$ and the  corresponding
small-scale magnetic field is given by 
${\bm b}^{pq} = \bm{\nabla} \times {\bm a}^{pq}$.
Note that at large values of $\Rm$ there will also be small-scale dynamo action
that will lead to spurious time-dependencies of $\meanEMF$.
However, since neither the test fields nor $\alpha_{ij}$ or $\eta_{ij}$
depend on time, i.e.\
\begin{equation}
\overline{\mathcal{E}}_i^{pq}(z,t) = \overline{\mathcal{E}}_{0i}^{pq}(z,t) +
                  \alpha_{ij} {\overline B_j}^{pq}(z) +
                  \eta_{ijk} {\overline B_{j,k}}^{pq}(z) \, ,
\label{emfpq}
\end{equation}
such time dependence must be entirely due to $\meanEMF_0^{pq}(z,t)$
and can be eliminated by time averaging.

In the following we shall, to begin with, use $k=k_1$, 
the wavenumber corresponding to the
box size, to study the dependence of $\alpha_{ij}$ and 
$\eta_{ij}$ on shear and magnetic Reynolds number.
We shall discuss the dependence of the $\alpha_{ij}$ and
$\eta_{ij}$ tensors on $k$ in Section \ref{scaledep}.
Note that the usual approach of using uniform applied fields for
calculating $\alpha$ \citep[e.g.,][]{Cou06} corresponds to a 
special case of the test field method for $k=0$.
However, dynamos generate large-scale fields with non-zero $k$,
so it is important to relax this restriction.
It is then also important to calculate $\eta_{ij}$. The test 
field method allows the simultaneous calculation of all the components of
the $\alpha_{ij}$ and $\eta_{ij}$ tensors for arbitrary values of $k$.

The test field method has recently been criticised by \cite{CH08} on the
grounds that the test fields are arbitrary predetermined mean fields.
They argue that the resulting turbulent transport coefficients
will only be approximations to the true values unless the test fields
are close to the actual mean fields
-- a criticism equally applicable to other methods using arbitrary
uniform fields.
This concern has been addressed by \cite{TB08}, who argue that
Eq.~(\ref{da_dt}) can instead be applied to any mean field.
This statement has been numerically verified in three cases
that we describe below.

Firstly, the test field method correctly reproduces a vanishing growth
rate in saturated nonlinear cases~\citep{BRRS08}.
Secondly, in the time-dependent case, the test field method
correctly reproduces also a non-vanishing growth rate, but
in that case it is no longer permissible to express $\meanEMF$ in terms of
a multiplication of turbulent transport coefficients with the mean field
and its spatial derivatives.
One must therefore write Eq.~(\ref{emf}) as a convolution in time
\citep{HB08}.
Finally, the success of the test field method becomes particularly clear
when it is applied to a passive vector field that obeys a separate 
induction equation with a velocity field from a saturated dynamo \citep{TB08}.
This question was originally posed by \cite{CT08}.
In particular for the Roberts flow with a mean field of Beltrami type,
e.g.\ one that is proportional to $(\cos k_1z,\sin k_1z, 0)$, 
the $\alpha_{ij}$ tensor is anisotropic and has an additional component
proportional to $\meanB_i\meanB_j$ that tends
to quench the components of the isotropic part of $\alpha_{ij}$.
The fastest growing passive vector field is then
proportional to $(\sin k_1z,-\cos k_1z, 0)$.
This result has been confirmed both numerically and
using weakly nonlinear theory~\citep{TB08}.

In the following we ignore the complications involving time-dependent
mean fields and restrict ourselves
to transport coefficients that apply strictly speaking only to the
time-independent or marginally excited case.
For our numerical simulations we use the \textsc{Pencil Code}\footnote{
\url{http://www.nordita.org/software/pencil-code}},
where the test field algorithm has already been implemented.
All our numerical simulations are performed in a periodic cubic box. 
The forcing scale is chosen to have the wavenumber $\kf/k_1=5$.
This gives enough scale separation for a large-scale field to develop
(Haugen et al.\ 2004), and is still not too big to reduce the resulting
Reynolds numbers too much.
We use units such that $c_{\rm s}=k_1=\rho_0=\mu_0=1$
and arrange the forcing amplitude
so that the Mach number is around $0.1$.
For the magnetic Prandtl number we choose $\Pm=\nu/\eta=1$, where $\nu$ lies
in the range  $3\times10^{-3}$ to $2\times10^{-4}$ (in units of $\cs/k_1$).
We choose the shear $S$ such that the parameter
\begin{equation}
\mbox{Sh}\equiv S/(u_{\rm rms} k_{\rm f})
\end{equation}
takes values in the range $-0.02$ to $-0.9$, where $\urms$ is the 
root-mean-square velocity. 
All the runs are started with uniform density, $\rho=\rho_0$, 
${\bm U}=0$, and ${\bm a}^{pq}=0$.
Depending upon the parameters of a particular run we use up to 
$256^3$ grid points. For each test field ${\overline {\bm B}}^{pq}$ 
we calculate the time averages of
the $\alpha_{ij}$ and $\eta_{ij}$ tensors over time intervals over which 
$u_{\rm rms}$ is statistically stationary. In runs 
with higher magnetic Reynolds numbers,
\begin{equation}
\Rm \equiv u_{\rm rms}/(\eta k_{\rm f}),
\end{equation}
we obtain an exponential growth of
small-scale magnetic field (see below).
We interpret this as being associated with the $\meanEMF_0^{pq}(z,t)$ term.
This often gives rise to large fluctuations in all components of 
the $\alpha_{ij}$ and $\eta_{ij}$ tensors
at late times. In such cases we confine our calculations of the 
time averages to time intervals over which
${\bm b}^{pq}_{\rm rms}$ does not exceed ${\overline {\bm B}}^{pq}$
by more than a factor of about 20.
Up until this point, the components of the 
$\alpha_{ij}$ and $\eta_{ij}$ tensors
show only their intrinsic fluctuations, but at later times these will
be swamped by additional contributions from the small-scale dynamo
that grows exponentially in time.
If necessary, we repeat our 
calculations over several independent realizations by resetting ${\bm a}^{pq}=0$
at regular time intervals. 

\section{Results}
\label{results}
Our principal aim in this paper is to study the effects of varying shear~($\mbox{Sh}$)
and magnetic Reynolds number~($\Rm$) on the components
of the $\alpha_{ij}$ and $\eta_{ij}$ tensors. 
In the subsection below we summarize 
our results concerning the different components of these tensors.

\subsection{Diagonal components of the transport tensors}

The isotropic parts of the $\alpha_{ij}$ and $\eta_{ij}$
tensors are respectively characterised as
\begin{equation}
\alpha = {\textstyle\frac{1}{2}}\langle \alpha_{11}+\alpha_{22} \rangle, ~~~
\eta_{\rm t} = {\textstyle\frac{1}{2}} \langle \eta_{11}+\eta_{22} \rangle,
\label{alpeta}
\end{equation}
where $\langle \cdot \rangle$ denotes an average
taken over $z$ and $t$. We normalize these quantities by  
$\alpha_0=-{\textstyle\frac{1}{3}}u_{\rm rms}$ and 
$\eta_{{\rm t}0}={\textstyle\frac{1}{3}}u_{\rm rms}/k_{\rm f}$
which are their respective expressions obtained using 
the First Order Smoothing Approximation~(FOSA) for $\Rm\ll1$,
which has previously been confirmed with the test field method in simulations of
helical turbulence without shear~\citep{SBS07}.

\begin{figure*}\begin{center}
\includegraphics[width=\columnwidth]{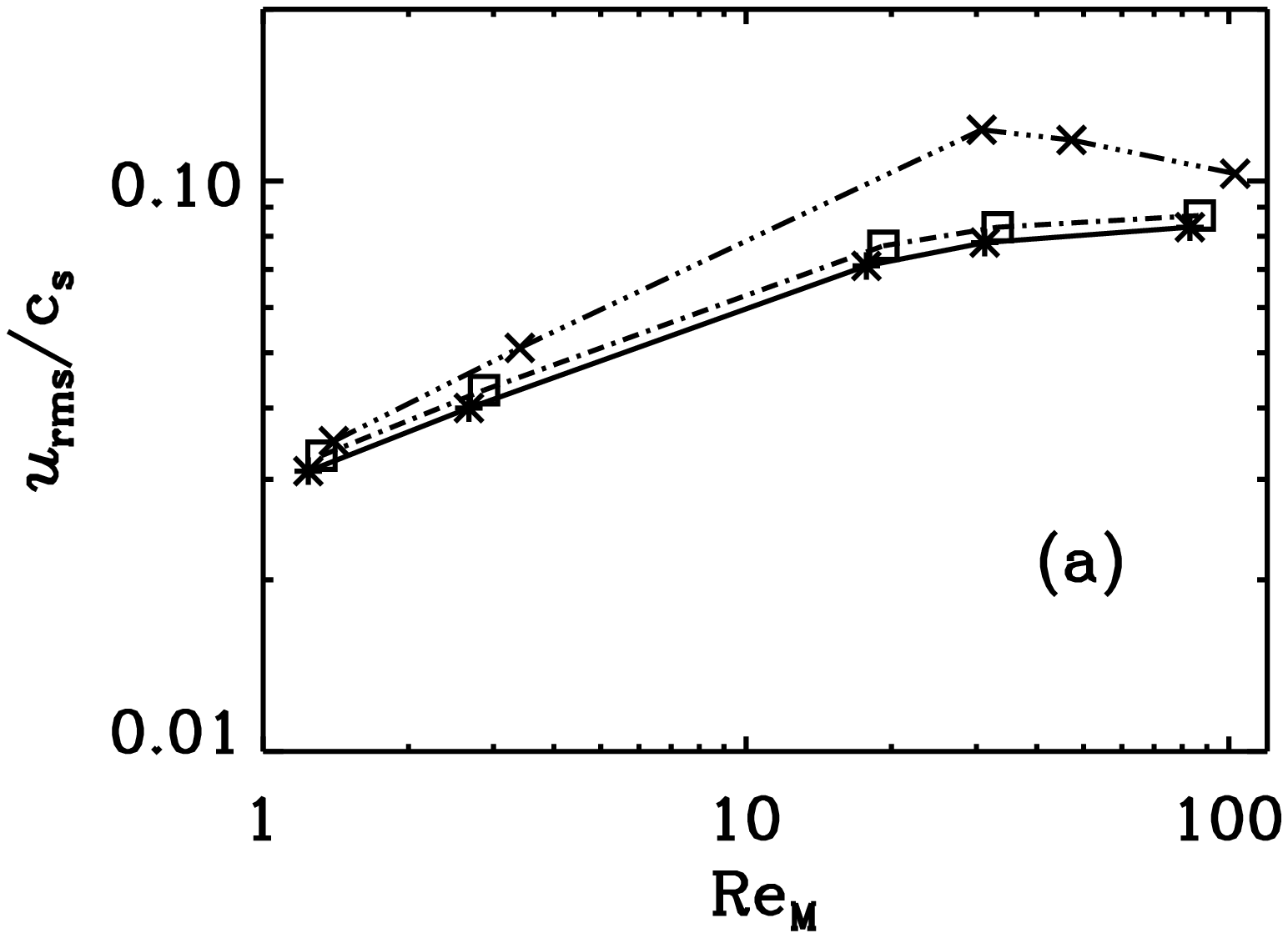}
\includegraphics[width=\columnwidth]{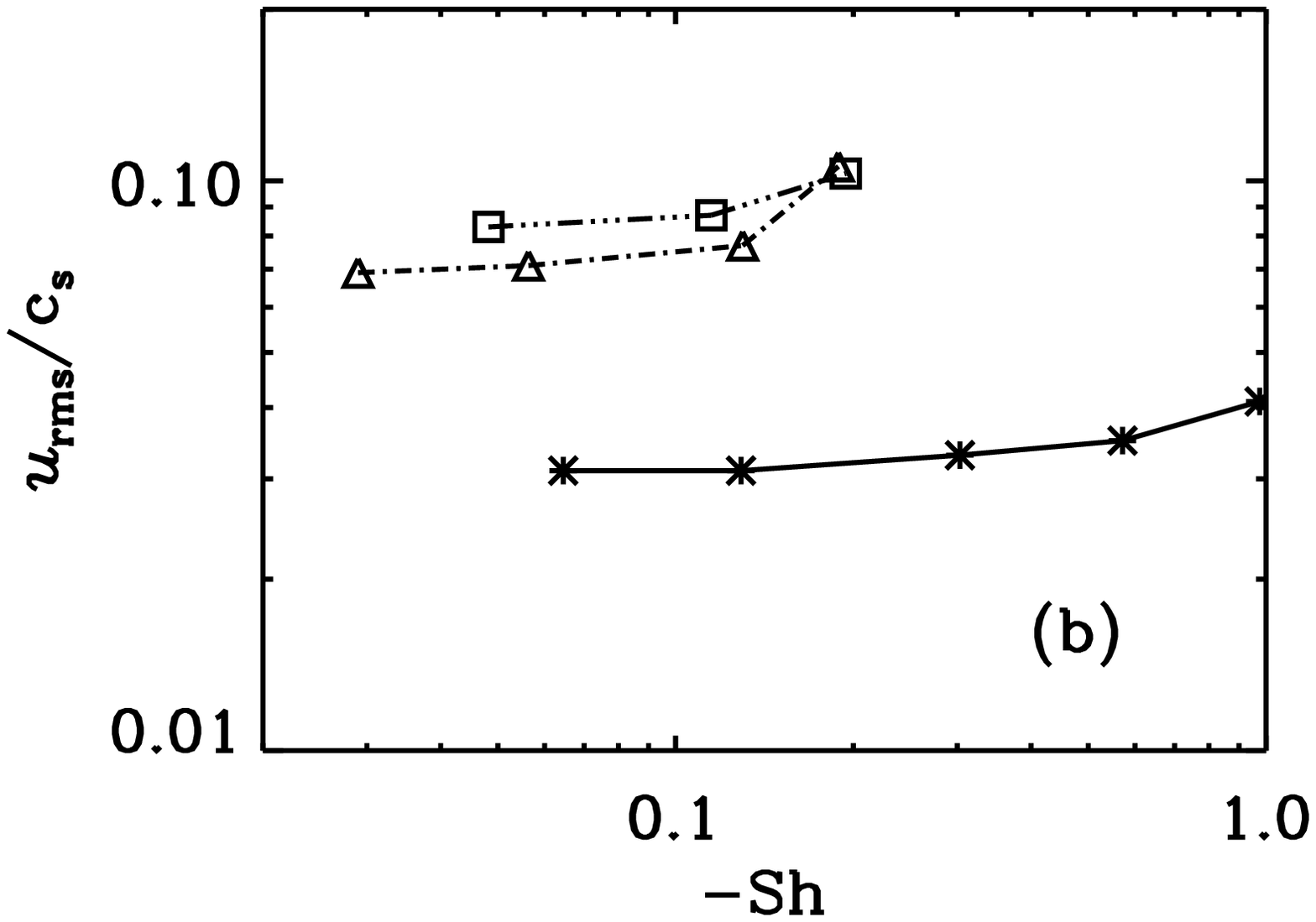}
\end{center}
\caption[]{Root-mean-square velocity $u_{\rm rms}$ normalized by
the speed of sound $c_{\rm s}$ as: (a) a function
of $\Rm$ for different values of $\mbox{Sh}$: $\mbox{Sh}\sim 0.07 (\ast), 
0.2  (\triangle), \mbox{and} 0.3 (\Box)$ respectively and 
(b) a function of the shear parameter
$\mbox{Sh}$ for different value of $\Rm$: 
$\Rm \sim 1$  ($\ast$), $\Rm \sim 20$ ($\triangle$), and  
$\Rm \sim 72$ ($\square$).
}\label{urms}
\end{figure*}

Since $u_{\rm rms}$ enters the normalization of $\alpha_{ij}$ and $\eta_{ij}$,
it is useful to first look at how it changes as a function of 
$\Rm$~(Fig.~\ref{urms}a) and $\mbox{Sh}$~(Fig.~\ref{urms}b).
As can be seen, $u_{\rm rms}$ 
increases as a function of $\Rm$ for small $\Rm \sim 1$ and then reaches
a plateau for high $\Rm$. On the other hand, $u_{\rm rms}$ 
is almost a constant as a function of shear except for high $\mbox{Sh}$,
where we observe excitation of the vorticity dynamo discussed further in 
Section~\ref{vortdyn}.

\begin{figure*}\begin{center}
\includegraphics[width=\textwidth]{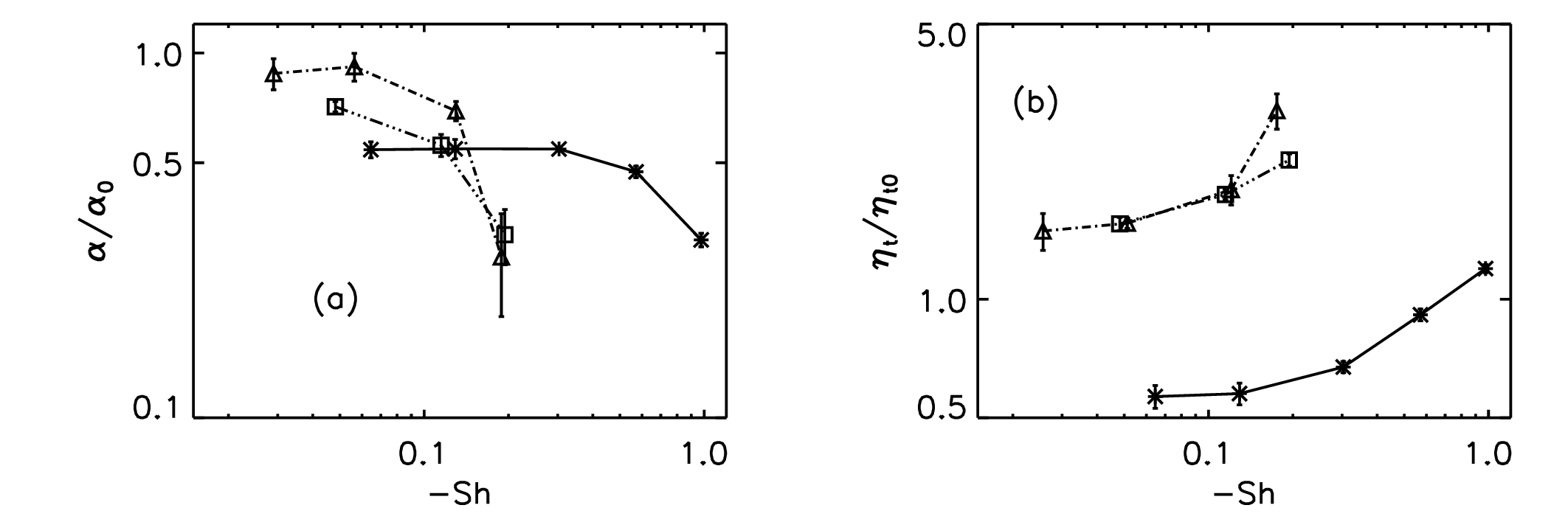}
\end{center}
\caption[]{Turbulent transport coefficients (a) $\alpha/\alpha_0$  and 
(b) $\eta_{\rm t}/\eta_{{\rm t}0}$  as functions of shear 
parameter $\mbox{Sh}$ for different values 
of $\Rm$: $\Rm \sim 1$  ($\ast$), $\Rm \sim 20$ ($\triangle$), and  
$\Rm \sim 72$ ($\square$).
}\label{pvs_shear}
\end{figure*}

\begin{figure*}\begin{center}
\includegraphics[width=\textwidth]{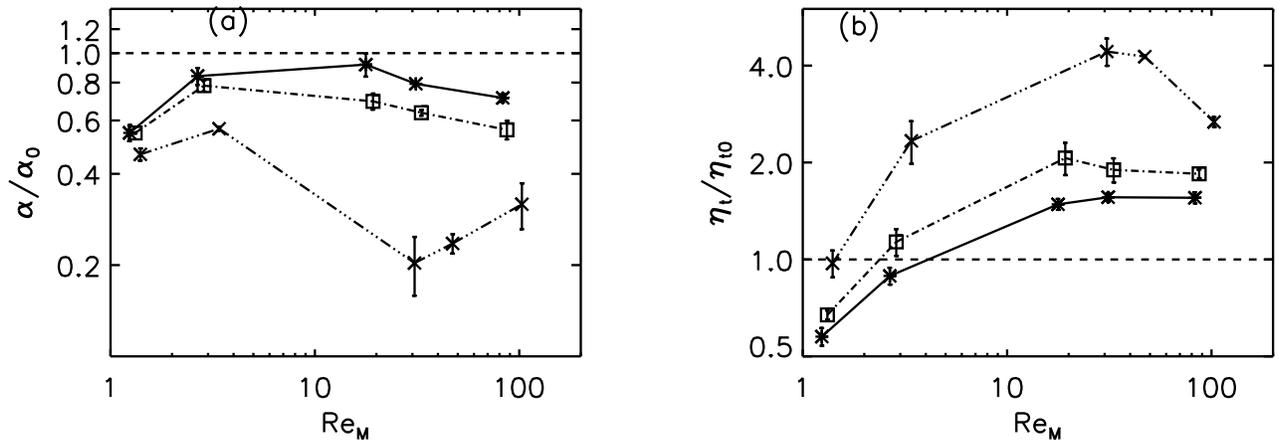}
\end{center}
\caption[]{
(a) $\alpha/\alpha_0$  and (b) $\eta_{\rm t}/\eta_{{\rm t}0}$  as functions of $\Rm$,
 for different values of $\mbox{Sh}$: $\mbox{Sh}\sim 0.07 (\ast), 
0.2  (\triangle), \mbox{and}  0.3 (\Box)$ respectively.
Horizontal dashed lines
at $\alpha/\alpha_0 = 1$ and $\eta_{\rm t}/\eta_{{\rm t}0}=1$ are added to 
facilitate comparison.} 
\label{pvs_rm}
\end{figure*}

In Fig.~\ref{pvs_shear} we show $\alpha$ as a function of shear for three
different values of the magnetic Reynolds number $\Rm \approx 1, 20$ 
and $72$. Figure~\ref{pvs_shear}b shows the corresponding
results for $\eta_{\rm t}$. We note that for small shear  
the turbulent transport coefficients are close to their values for 
zero shear.
As $|\mbox{Sh}|$ increases, $\alpha$ decreases and $\eta_{\rm t}$ increases 
up to four times $\eta_{\rm t0}$.
As can be seen there is a clear reduction (quenching) of $\alpha$ with
increasing shear in all these cases.
In order to examine the possible convergence of the results with $\Rm$ we plot
$\alpha$ and $\eta_{\rm t}$
as functions of $\Rm$ for three different values of the shear parameter
$\mbox{Sh}$, see Fig.~\ref{pvs_rm}.
For small $\Rm \sim 1$ we observe an increase in both $\alpha$ and
$\eta_{\rm t}$ with $\Rm$.
A similar initial increase of $\alpha$ and $\eta_{\rm t}$ was also seen in
earlier simulations of helical turbulence without shear~\citep{SBS07} and in  
non-helical shear flow turbulence \citep{BRRK08}.
For higher values of $\Rm$ and $\mbox{Sh}$, both $\alpha$ and $\eta_{\rm t}$
show large variations.
In these kinematic simulations, however, we expect
them to tend to constant values asymptotically at high $\Rm$ \citep{SBS07}.

\begin{figure*}\begin{center}
\includegraphics[width=\textwidth]{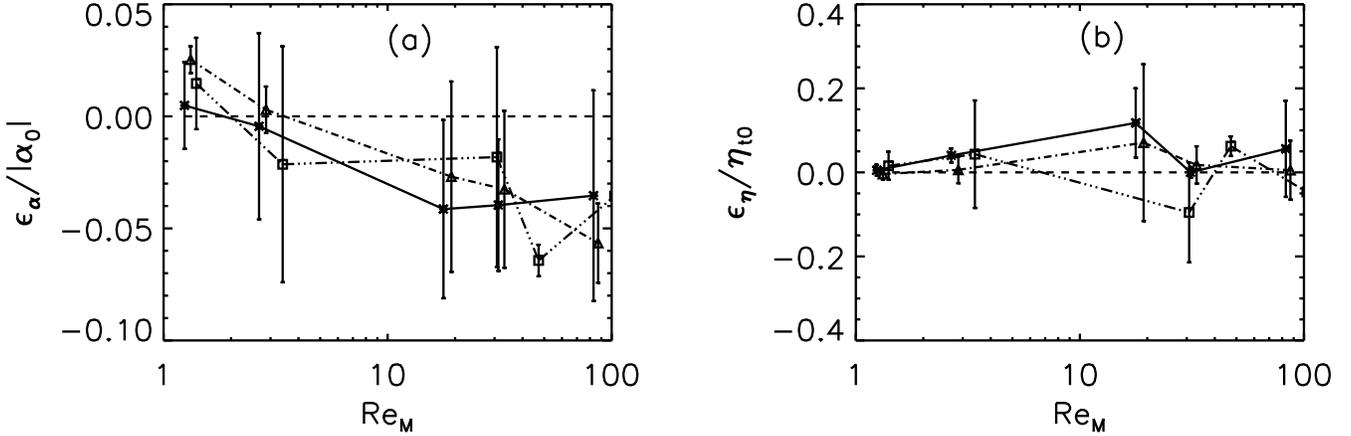}
\end{center}
\caption[]{Normalized 
(a) $\epsilon_{\alpha}$ and (b) $\epsilon_{\eta})$,
as defined in Eq.~(\ref{epsilon}),
as functions of $\Rm$ for constant value of the shear parameter
$\mbox{Sh}$: $\mbox{Sh}\sim 0.07 (\ast), 0.2  (\triangle), \mbox{and} \ 
0.3 (\Box)$ respectively. 
Horizontal dashed lines are added to facilitate comparison.}
\label{eps}
\end{figure*}

In an earlier study of shear flow turbulence with non-helical forcing by
\cite{BRRK08}, the diagonal components  of $\eta_{ij}$ were
found to be the same.
Obviously, in the absence of helicity  
all the components of $\alpha_{ij}$ are zero.
It turns out that in the presence of helicity the two diagonal components
of $\eta_{ij}$ are still the same, but those of $\alpha_{ij}$ are now
non-zero and not equal to each other.
This is best shown by considering the quantities
\begin{equation}
\epsilon_{\alpha} = {\textstyle\frac{1}{2}}\langle \alpha_{11}-\alpha_{22} \rangle, ~~~
\epsilon_{\eta} = {\textstyle\frac{1}{2}} \langle \eta_{11}-\eta_{22} \rangle . 
\label{epsilon}
\end{equation} 
The results are shown in Fig.~\ref{eps}.
Note that especially for large values of $\Rm$ the values of
$\epsilon_{\alpha}$ are predominantly negative.
Since both $\alpha_{11}$ and $\alpha_{22}$ are negative,
this means that $|\alpha_{11}|$ is larger than $|\alpha_{22}|$,
although the relative difference is only of the order of at most 5 per cent.

\begin{figure*}\begin{center}
\includegraphics[width=\textwidth]{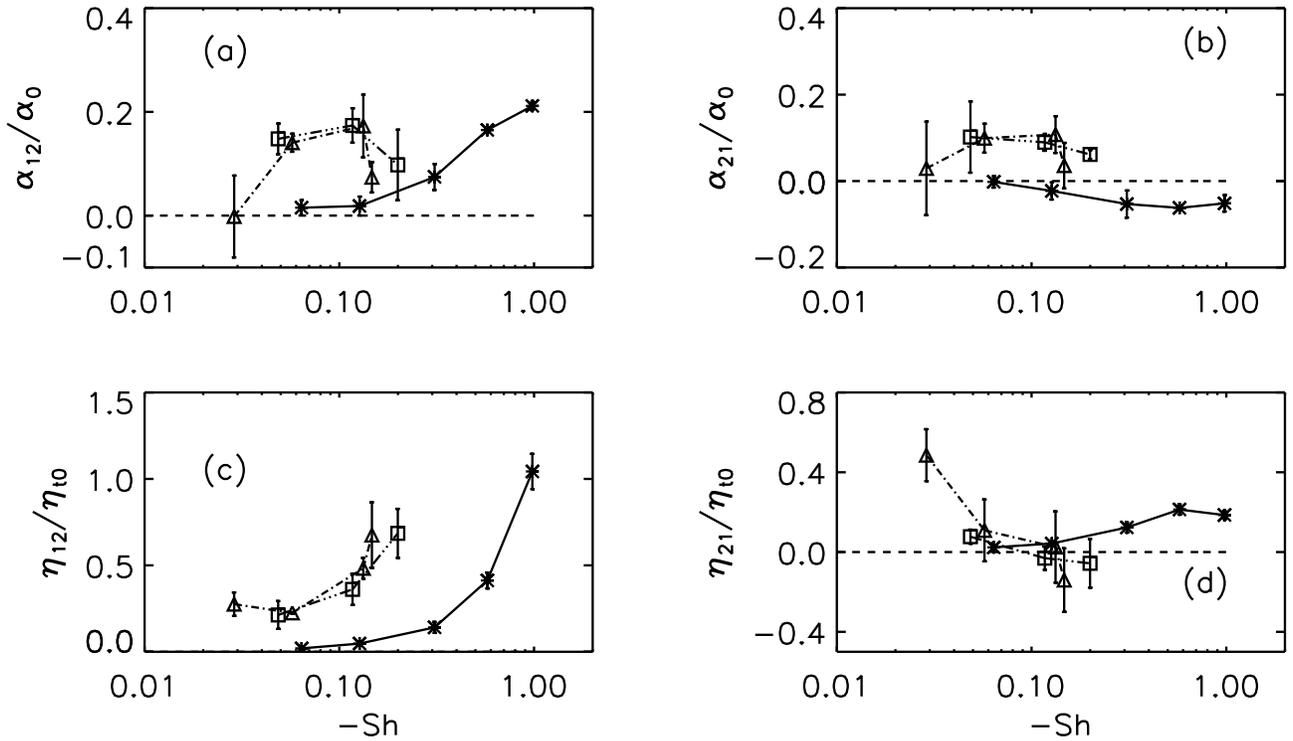}
\end{center}
\caption[]{Normalized off-diagonal components of
the $\alpha$ and $\eta$ tensors as functions of the shear
parameter $\mbox{Sh}$  for different values
of $\Rm$: $\Rm \sim 1$  ($\ast$), $\Rm \sim 20$ ($\triangle$), and
$\Rm \sim 72$ ($\square$).
Horizontal dashed lines are added to facilitate comparison.}
\label{podd_vsS}
\end{figure*}
\begin{figure*}\begin{center}
\includegraphics[width=\textwidth]{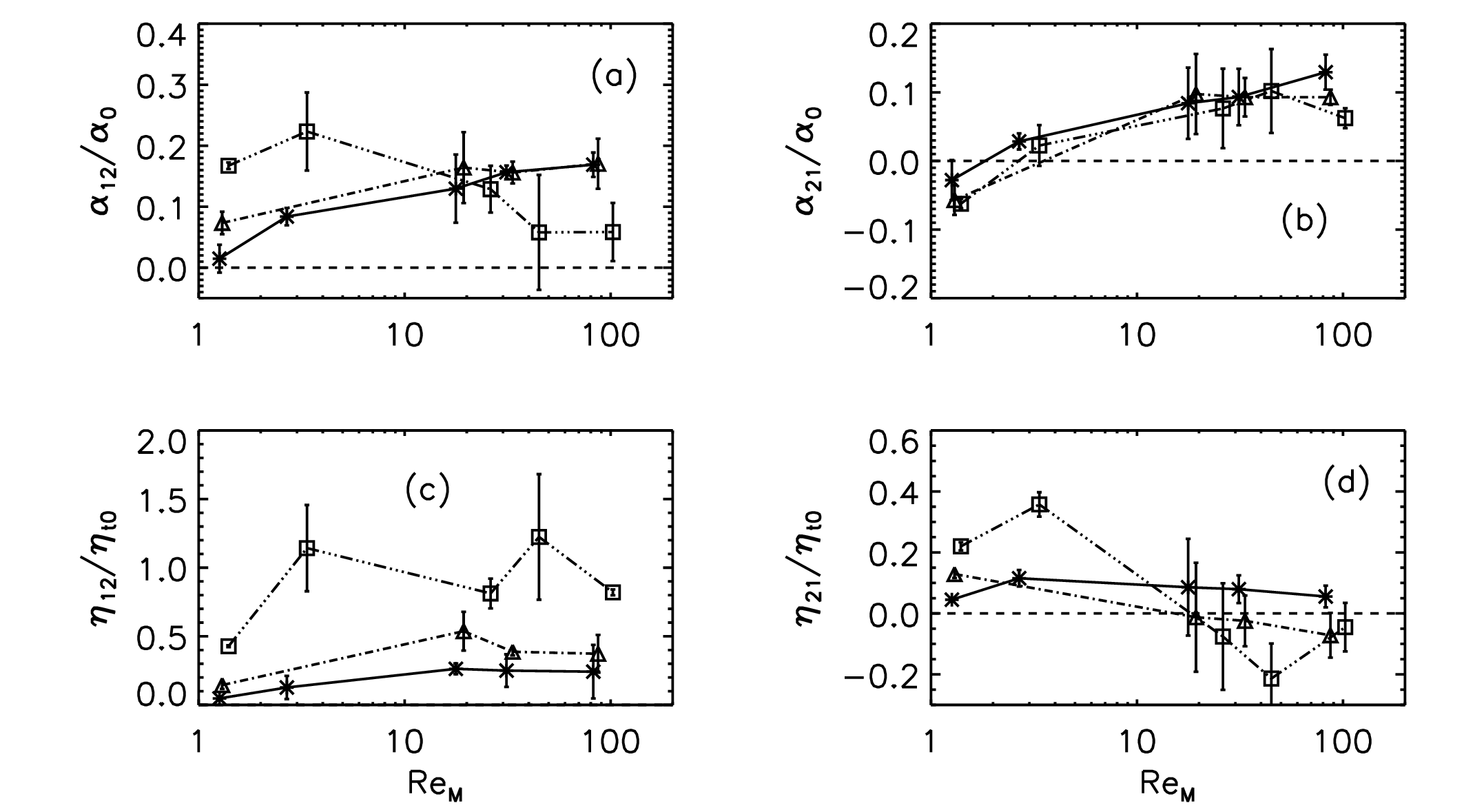}\\
\end{center}
\caption[]{Normalized off-diagonal components of the $\alpha$ and 
$\eta$ tensors as functions of $\Rm$ for constant values of 
$\mbox{Sh}$: $\mbox{Sh}\sim 0.07~(\ast),~ 0.2~(\triangle),~ \mbox{and} \ 0.3 ~(\Box)$. 
Horizontal dashed lines are added to facilitate comparison.}
\label{podd}
\end{figure*}

\subsection{Off-diagonal components}

\begin{figure*}\begin{center}
\includegraphics[width=\textwidth]{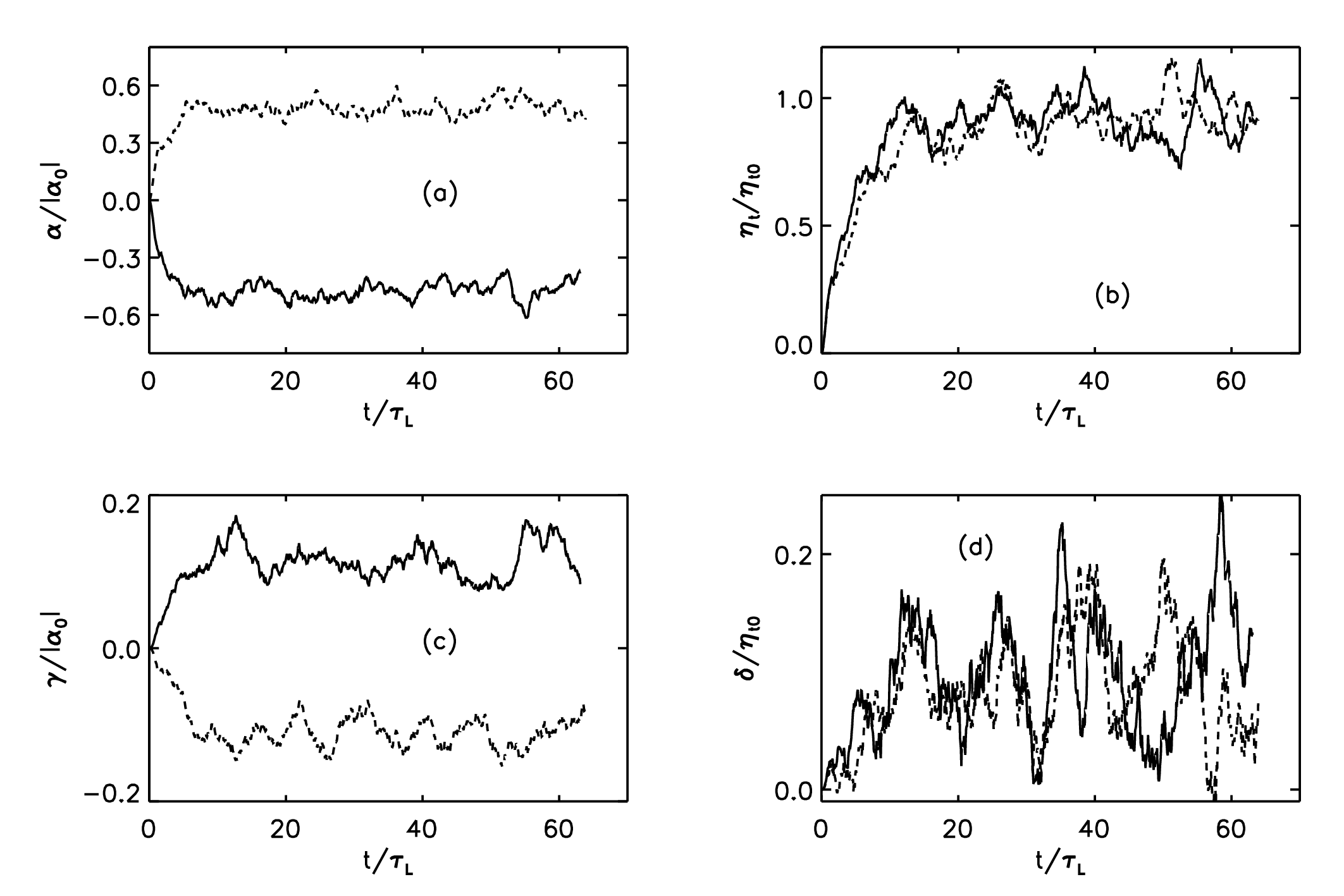}
\end{center}
\caption[]{Normalized time series of turbulent transport coefficients, 
(a) $\alpha/|\alpha_0|$, (b) $\eta_t/\eta_{t0}$, 
(c) $\gamma/|\alpha_0|$ and (d) $\delta/\eta_{t0}$, from two runs 
with exactly the
same parameters~($\Rm\sim 1.4$, $\mbox{Sh}=-0.5$), but different signs of 
helicity for the external force. 
Continuous and broken lines denote results from runs with positive and negative helicities, 
respectively.}
\label{paet}
\end{figure*}

We now consider the off--diagonal components of $\alpha_{ij}$ and $\eta_{ij}$.
The results are depicted in Figs.~\ref{podd_vsS} and \ref{podd}.
Of particular interest among these is the component $\eta_{21}$,
whose dependence on $\Sh$ and $\Rm$ is shown
in Figs.~\ref{podd_vsS}d and \ref{podd}d, respectively.
This component can indicate the possible presence of a 
shear--current dynamo that may operate when $\eta_{21}S/(\eta_{\rm T} k_1)^2>1$
\citep{RK03,RK04}. Here $\eta_{\rm T}=\eta_{\rm t}+\eta$ is the total
(sum of turbulent and microscopic) magnetic diffusivity.
In our case we have $S < 0$,
which implies that the necessary condition for the shear--current dynamo 
to operate is $\eta_{21} < 0$.  As can be seen from Fig.~\ref{podd}d,
for the range of $\Rm$ values considered here $\eta_{21}$ is positive for
small shear but becomes negative for strong shear and certain values of $\Rm$.
Earlier work of \cite{BRRK08} without helicity did indicate a similar
sign change, although only for larger $\Rm$. However, the error bars were so large that
this result could not be regarded as significant.
For the run with the strongest shear ($-\Sh\approx0.3$) and for $\Rm=40$
we now find $\eta_{21}$ to be more clearly negative, but for smaller
and larger values of $\Rm$ the results are again, at least within
error bars, compatible with zero.
Also, of course, the present results apply to the case with helicity
and are therefore not really comparable with those of \cite{BRRK08},
where the helicity is zero.
The antisymmetric contributions to the $\alpha_{ij}$ and $\eta_{ij}$ tensors
are characterised by the vectors
\begin{equation}
\gamma_k= -\onehalf \epsilon_{ijk}\alpha_{ij},\quad
\delta_k= -\onehalf \epsilon_{ijk}\eta_{ij}.
\end{equation}
Since our averages depend only on $z$, the $z$ components
of these tensors are irrelevant and therefore only the
$z$ components of the $\bm{\gamma}$ and $\bm{\delta}$ vectors
are of interest.
We denote those simply by $\gamma$ and $\delta$, with
\begin{equation}
\gamma = {\textstyle\frac{1}{2}}\langle \alpha_{21}-\alpha_{12} \rangle,\quad
\delta = {\textstyle\frac{1}{2}} \langle \eta_{21}-\eta_{12} \rangle.
\label{gd}
\end{equation} 
A time series of both quantities is shown in Fig.~\ref{paet} for
positive and negative signs of the kinetic helicity.
The results show that, in our case with $S<0$,
$\gamma$ is positive (negative) for positive (negative) kinetic helicity,
whilst $\delta$ has always the same sign.

\begin{figure*}\begin{center}
\includegraphics[width=\textwidth]{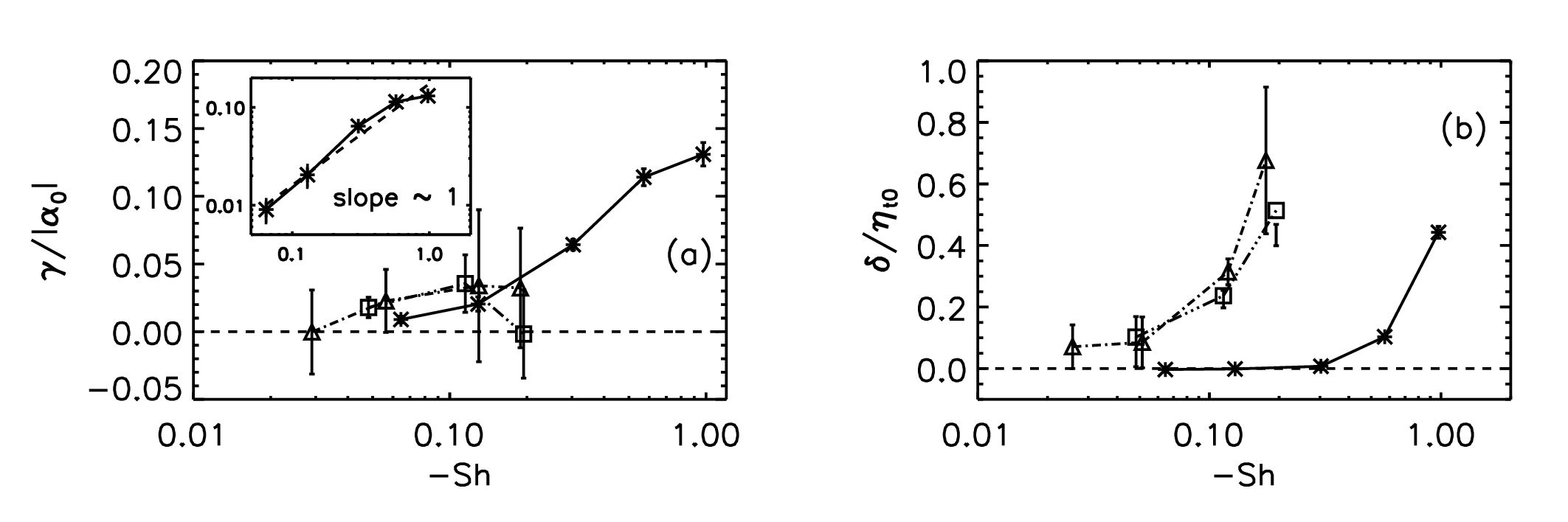}
\end{center}
\caption[]{Normalized 
(a) $\gamma$ and (b) $\delta$  
as functions of $\mbox{Sh}$ for
$\Rm \approx 1$  ($\ast$), $\Rm \approx 20$ ($\triangle$), and  
$\Rm \approx 72$ ($\square$).
Horizontal dashed lines are added to facilitate comparison.
The inset shows $\gamma/|\alpha_0|$ versus
$-\mbox{Sh}$ for $\Rm\sim1$.
}\label{gdvs_shear}
\end{figure*}

\begin{figure*}\begin{center}
\includegraphics[width=\textwidth]{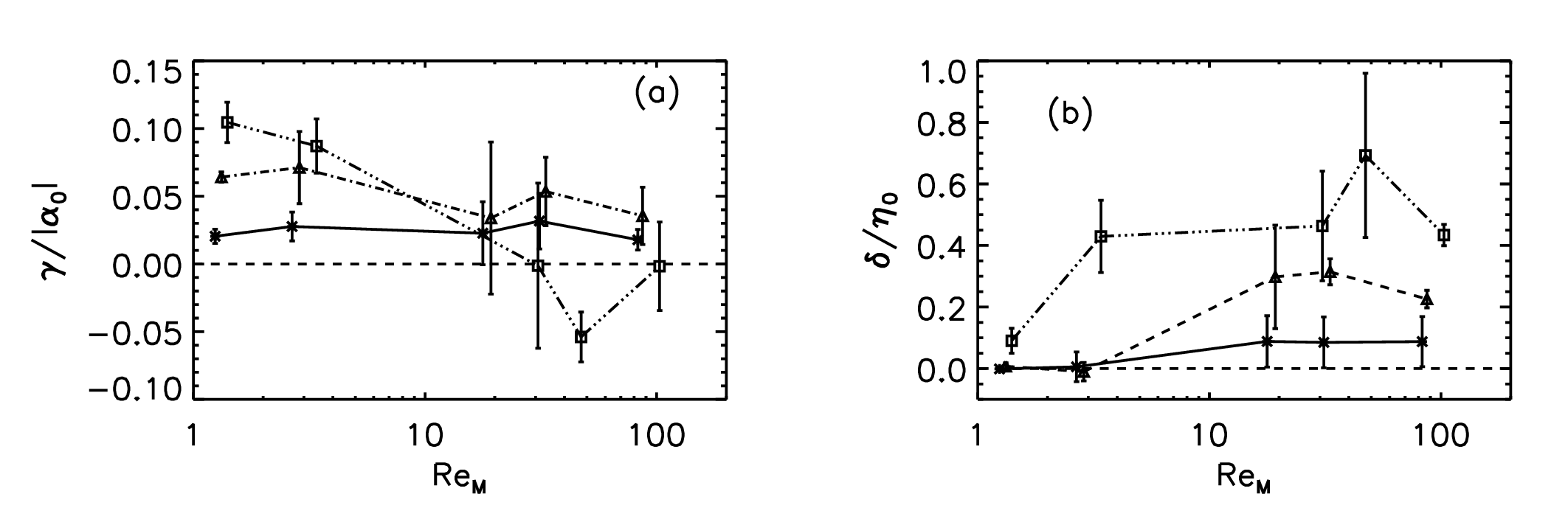}
\end{center}
\caption[]{Normalized (a) $\gamma$ and (b) $\delta$  
as functions of $\Rm$ for constant values of the shear parameter
$\mbox{Sh}$: $\mbox{Sh}\sim 0.07~(\ast),~ 0.2~ (\triangle), 
\mbox{and} \  0.3~ (\Box)$.
Horizontal dashed lines are added to facilitate comparison.}
\label{gamma-delta}
\end{figure*}

Some idea about the functional forms of $\gamma$ and $\delta$ can be
obtained from symmetry considerations.
The vectors $\bm{\gamma}$ and $\bm{\delta}$ enter the electromotive force
thus
\begin{equation}
\meanEMF=...+\bm\gamma\times\meanBB-\mu_0\bm\delta\times\meanJJ,
\end{equation} 
so we see that $\bm\gamma$ must be a polar vector and $\bm\delta$ must be
an axial vector.
Using the shear flow $\overline{\bm U}^{S}$, the only axial vector that
can be constructed is the mean vorticity,
$\meanWW={\bm\nabla}\times\overline{\bm U}^{S}$, so we expect
$\bm\delta$ to have a component that is proportional to $\meanWW$.
Likewise, since $\bm{\gamma}$ is a polar vector which points in the
direction of $\meanWW$, the two must be related via a pseudoscalar.
In the present case the only pseudoscalar available is the kinetic
helicity, $\overline{{\bm\omega}\cdot{\bm u}}$.
Based on these symmetry arguments we write
\begin{equation} 
\bm{\gamma}=C_\gamma\tau^2\overline{{\bm\omega}\cdot{\bm u}}\,\meanWW,\quad
\bm{\delta}=C_\delta\tau^2\overline{{\bm u}^2}\,\meanWW,
\label{eqgda}
\end{equation}
where we have introduced two non-dimensional quantities,
$C_\gamma$ and $C_\delta$ 
which could be either positive or negative,
and $\tau$ is a correlation time that we approximate here by
\begin{equation}
\tau=(\urms\kf)^{-1}.
\end{equation}
In the present case, $\overline{{\bm\omega}\cdot{\bm u}}$ itself is
a negative multiple of $\alpha$.
We therefore expect $\bm{\gamma}$ to have a component proportional to
$\alpha\meanWW$, multiplied by the correlation time $\tau$.
By similar arguments we expect $\bm\delta$ to have a component proportional
to $\etat\meanWW$, multiplied by $\tau$.
Based on these arguments we can also write
\begin{equation} 
\bm{\gamma}=\tilde{C}_\gamma\,\alpha\meanWW\tau,\quad
\bm{\delta}=\tilde{C}_\delta\,\etat\meanWW\tau,
\label{eqgd}
\end{equation}
with new non-dimensional quantities,
$\tilde{C}_\gamma\approx C_\gamma/3$ and $\tilde{C}_\delta\approx C_\delta/3$,
that we expect to be of order unity.
Our simulations confirm this reasoning and show that both coefficients are of
order unity with positive $\tilde{C}_\gamma$ and negative $\tilde{C}_\delta$
(e.g., $\tilde{C}_\gamma= 0.5$ and $\tilde{C}_\delta= 0.25$ for the
run shown in Fig.~\ref{gdvs_shear})
for runs with small $\Rm$ and $\mbox{Sh}$;
see Figs.~\ref{gdvs_shear} and \ref{gamma-delta}, respectively.
The sign in Eq.~\ref{eqgd} has been chosen a posteriori from our simulations. 
For high shear and Reynolds number  this simple
reasoning is no longer accurate and, at least in one case
(Fig~\ref{gamma-delta}a), $C_\gamma$ even becomes negative. 
For lower
values of $\Sh$ we find an almost constant negative $\gamma$ of the
order of $0.05\alpha_0$ (or less).
This result is fairly independent of $\Rm$. 
The moduli of $\gamma$ and $\delta$ increase with increasing
shear parameter, and for larger values of $\Rm$ both quantities
may approach an asymptotic value. For small $\Rm$ we observe 
a roughly linear increase of $\gamma$ with shear (see inset of
Fig~\ref{gdvs_shear}a), further verifying Eq.~(\ref{eqgda}).

The coefficient $\gamma$ can be interpreted as turbulent pumping,
i.e.\ advection of the magnetic field by means other than the
mean velocity field.
In strongly stratified convection, turbulent pumping has been seen to be
directed from higher to lower turbulence intensity
\citep{TBCT98,TBCT01,OSBR02,KKOS06},
which is usually in the downward direction.
Thus, turbulent pumping is likely
to play an important role in convection zones of stars where it can
overcome the buoyancy of the magnetic field. In the present case where
the turbulence is homogeneous, however, stratification does not play
a role and the pumping is just due to the combined action of shear and
helical turbulence.

\subsection{Scale-dependence}
\label{scaledep}
\begin{figure*}\begin{center}
\includegraphics[width=\textwidth]{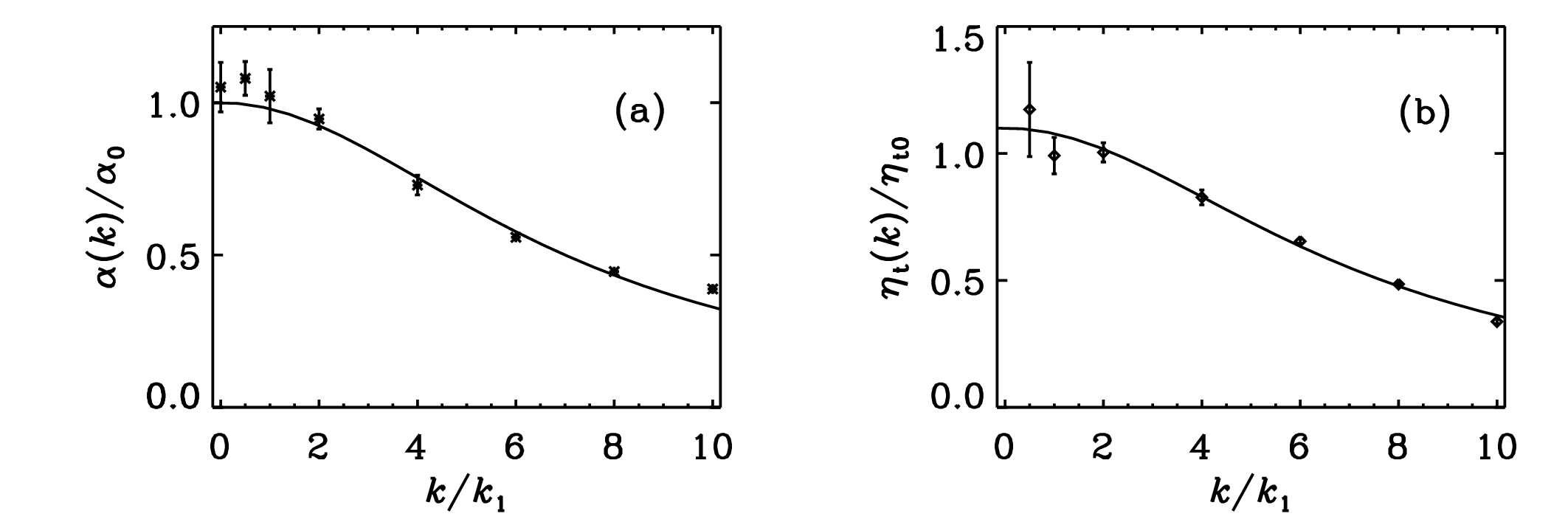}
\end{center}
\caption[]{(a) $\alpha(k)/\alpha_0$ and
(b) $\eta_t(k)/\eta_{\rm t0}$ as a function of the wavenumber $k$ of the
large-scale magnetic field for $\Rm\approx3$.
The solid lines show the Lorentzian fits.}
\label{k-depen}
\end{figure*}

So far we have confined the calculations of the $\alpha_{ij}$ and $\eta_{ij}$
tensors to test fields whose characteristic length scale is the largest scale
in the domain, i.e., test fields of the form $\sin kz$ or $\cos kz$ with 
$k=k_1$, where $k_1=2\pi/L$ and $L$ is the box size of our simulations. 
We have also done similar calculations for other values of $k$, which 
correspond to test fields with different characteristic length scales.
In Fig.~\ref{k-depen} we show the dependence of
$\alpha$ and $\eta_t$  on $k$ for $\Rm\approx3$.
The decrease
can be modelled by a Lorentzian peaked at $k=0$ (Fig.~\ref{k-depen}). 
This result is not surprising, because Lorentzian fits have been
obtained earlier also in the absence of shear \citep{BRS08}.

\begin{figure*}\begin{center}
\includegraphics[width=\columnwidth]{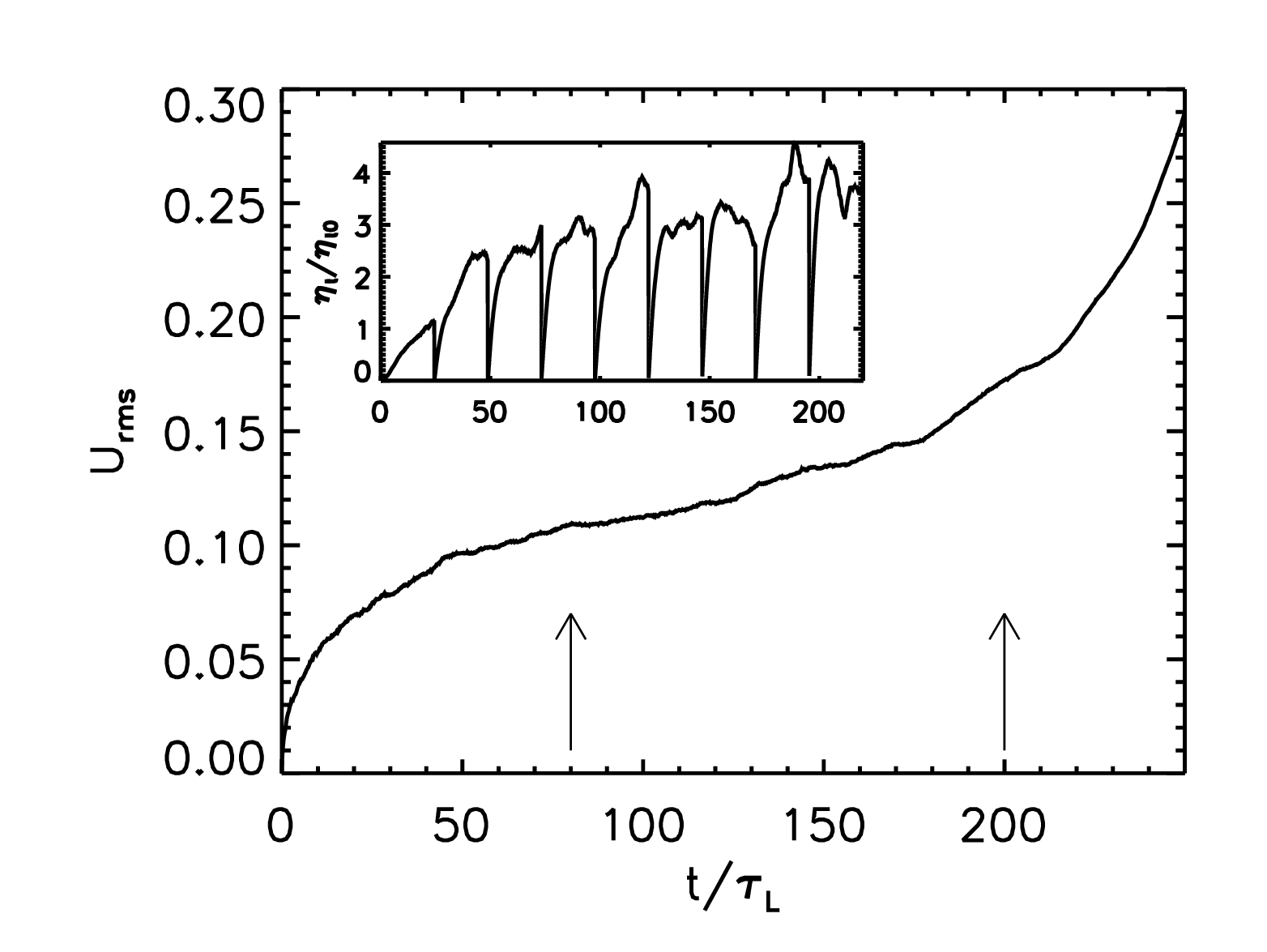}
\includegraphics[width=\columnwidth]{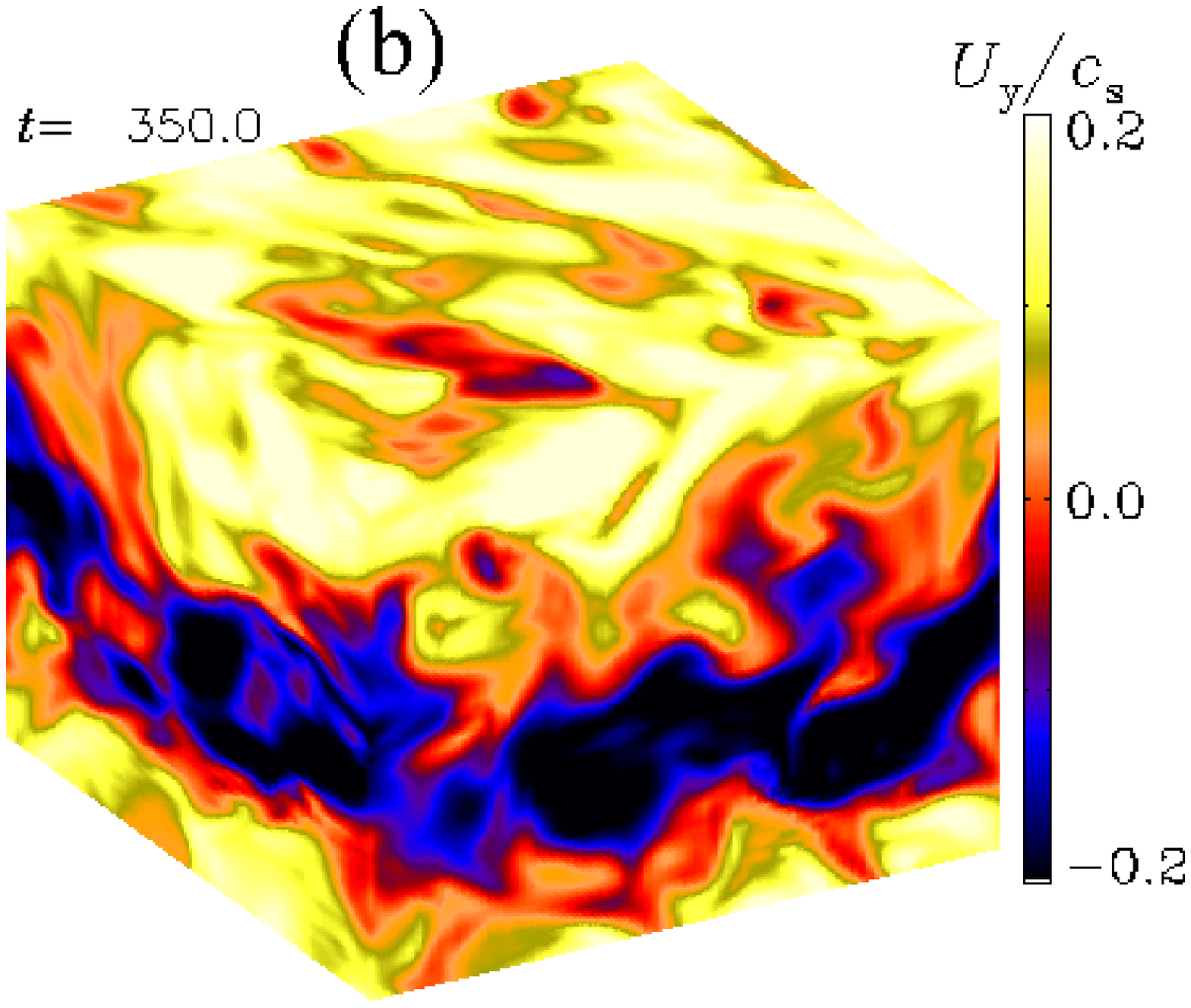}
\end{center}
\caption[]{
Panel (a) shows the root-mean-square values of the velocity for 
one of our runs (with ${\rm Re}\approx45$ and ${\rm Sh}\approx-0.3$)
which exhibits a vorticity dynamo. The inset shows the time-series for
$\eta_{\rm t}$.
Panel (b) shows a pronounced pattern in $U_y$ on the periphery of 
the box at the end of the same run. Similar structures are
also found in the $x$ component of the velocity. 
}
\label{Uy_box}
\end{figure*}

\subsection{Effects from the vorticity dynamo}
\label{vortdyn}
In the runs with relatively high values of shear and
Reynolds number, the root-mean-square velocity, $U_{\rm rms}$,
which initially reaches a steady state, shows an exponential growth at 
late times, as can be seen in Fig.~\ref{Uy_box}a. 
Similar behaviour is seen for
turbulent transport coefficients, which also
show large fluctuations for later times. Large fluctuations
of the turbulent transport coefficients at later times are also seen
by \cite{SBS07}, but those fluctuations are more irregular
and have a different origin.
They are interpreted as being due to
the development of a small-scale dynamo at late times
and are not due to the more systematic increase
in the rms velocity, which in turn
is associated with the shear.
However, we emphasize that even before
${\bm b}^{pq}$ becomes dominated by this type of dynamo action,
the temporal and spatial fluctuations of
$\alpha_{ij}$ and $\eta_{ij}$ are of the order of $\alpha_0$ and
$\eta_{\rm t0}$, respectively.
This is true even for large values of $\Rm$; see \cite{BRRK08}, who have
argued that these fluctuations also contribute to large-scale dynamo action
via the incoherent $\alpha$ effect \citep{VB97,P07}.
For example, the time-series of $\eta_t$ shown in the inset of Fig.~\ref{Uy_box}
has a plateau even beyond the range over which $U_{\rm rms}$ is steady.

The vertical spikes in the inset of Fig.~\ref{Uy_box}
come from resetting $\bm{a}^{pq}$ to zero
in regular time intervals; see the end of Sect.~\ref{TheModel}.
Note, however, that the mean values and also the upper envelope
trace the evolution of $U_{\rm rms}$ reasonably well,
including the increased rise after $t=300$.
The late time behaviour is accompanied by the formation of large-scale
vortical structures, as seen in Fig.~\ref{Uy_box}b,
which is a signature of the vorticity dynamo proposed by~\cite{EKR03},
see also \citep{YHSKRICM07,YHRSKRCM08}. A detailed numerical study of the
vorticity dynamo has been performed by \cite{KMB08}.

The presence of the vorticity dynamo and the resulting
systematic variation in $U_{\rm rms}$ as well as
the turbulent transport coefficients
for late times limit the lengths of time 
over which average values of the components of
the $\alpha_{ij}$ and $\eta_{ij}$ tensors
can be calculated. In the above case this interval lies between
the two arrows in Fig~\ref{Uy_box}a. 
This limits the range of $\mbox{Sh}$ and $\Rm$ that we have
probed and explains the larger
error bars shown for example in Fig.~\ref{podd},
and why they cannot be reduced by simply
running our simulations for longer times.
This problem would be avoided in the presence of magnetic fields, because
then the resulting Lorentz force would quench the vorticity dynamo
\citep{KB08}.
This is however beyond the scope of this paper.

\section{Conclusions}
We have studied the effects of varying shear and magnetic Reynolds 
number on the turbulent transport coefficients in the presence of helicity in 
the kinematic limit. We have shown that for fixed $\Rm$, $\alpha$ is 
reduced (quenched) with increasing shear. Despite the differences in the 
assumptions made, this quenching is qualitatively similar to 
the recent results obtained by \cite{LK08}. 
To the best of our knowledge this is the first numerical study 
to show quenching of $\alpha$
as a function of shear in helical turbulence.
We find that $\eta_{\rm t}$ increases 
with increasing shear in the range of $\Rm$ 
values that we have considered here. A similar behaviour
for $\eta_{\rm t} $ was also seen in \cite{BRRK08} where the 
forcing was non-helical. 

We also compute the off--diagonal components of the
$\alpha_{ij}$ and $\eta_{ij}$ tensors.
The antisymmetric part of $\alpha_{ij}$ corresponds to a turbulent
pumping velocity $\bm{\gamma}$ in the direction perpendicular to the
plane of the shear flow.
It shows a roughly linear increase with shear for small magnetic
Reynolds number. We propose simple expressions for $\gamma$ 
and $\delta$ in Eq.~(\ref{eqgda}) which show reasonable agreement with our 
numerical results for small $\mbox{Sh}$ and $\Rm$.   
Our expression shows that for negative helicity, $\bm{\gamma}$  
points in the direction of the vorticity of the mean flow.
Regarding the component $\eta_{21}$ we find indications that, at least in
one or two cases, this component changes sign and becomes negative.
This could be of significance in connection with the shear--current effect.

We also find that all the turbulent transport 
coefficients depend on the wavenumber of the
mean flow in a Lorentzian fashion, just as in the case of non-shearing
turbulence \citep{BRS08}.
This means that the kinematic values of $\alpha$ and $\eta_{\rm t}$
for $k=k_1$ are close to the values obtained for $k\to0$.
This is not the case for certain non-turbulent flows such as
the Galloway-Proctor flow \citep{Cou08,RB08}.
In an earlier paper, \cite{Cou06} considered only the limiting case
$k=0$ for this flow.

Several aspects of the present investigations could
be of astrophysical relevance.
Turbulence in celestial bodies is helical and exhibits an $\alpha$ effect.
In addition, shear ($S$) can be an important ingredient 
in that the efficiency of large-scale dynamo action is determined by
the product of $\alpha$ and $S$.
However, as $S$ increases, $\alpha$ itself becomes quenched when $S$
becomes comparable with the inverse turnover time, i.e.\ $S\tau=O(1)$.
Furthermore, the turbulent diffusivity becomes enhanced, suppressing the
dynamo even further.
Finally, it is found that the combined action of helicity and shear
gives rise to a pumping velocity of mean magnetic field perpendicular
to the plane of the shear flow.
The existence of such a pumping velocity has not been emphasized before.
On the other hand, it is well known that $\alpha\Omega$ (or rather $\alpha S$)
dynamos can have travelling wave solutions~\citep{BBS01}.
When the product of $\alpha$ and $S$ is positive, these waves travel
in the positive $z$ direction, which agrees with the direction of pumping.
In the near-surface shear layer of the Sun this pumping would therefore
support the equatorward migration in that layer.

It is important to understand the quenching of turbulent 
transport coefficients in the presence of shear beyond the 
kinematic approximation.
In that case one needs to include the induction equation (in addition to
the test field equations) and incorporate the resulting Lorentz force
in the momentum equation  \citep{BRRS08}.
This would also help in alleviating problems of strong late-time fluctuations
arising from the vorticity dynamo, because the vorticity dynamo tends to
be suppressed by magnetic fields of equipartition strengths~\citep{KB08}.
Similarly, given their potential importance in allowing the
escape of magnetic helicity, the effects of open boundary conditions
also needs to be considered. These questions are under study and 
will be reported elsewhere.

\begin{acknowledgements}
  The authors acknowledge the hospitality of Nordita during the
  programme `Turbulence and Dynamos'.
  AB and PJK thank Astronomy Unit, 
 Queen Mary University of London, for hospitality. Computational resources were granted
  by CSC (Espoo, Finland), UKMHD, and QMUL HPC facilities purchased
  under the SRIF initiative.
  This work was supported by the the Leverhulme Trust (DM, RT),
  the Academy of Finland grant No.\ 121431 (PJK),
  and the Swedish Research Council (AB).
\end{acknowledgements}


\end{document}